\documentclass[twocolumn,showpacs,preprintnumbers,amsmath,amssymb]{revtex4}

\usepackage{graphicx}
\usepackage{dcolumn}
\usepackage{bm}

\begin{document}

\title{The 'Fast Exchange' model visualized with $^3$He confined in aerogel: \\
a Fermi liquid in contact with a Ferromagnetic solid}
\author{E. Collin, S. Triqueneaux, Yu.M. Bunkov and H. Godfrin}

\address{
Institut N\'eel
\\
CNRS et Universit\'e Joseph Fourier, \\
BP 166, 38042 Grenoble Cedex 9, France \\
}

\date{\today}

\begin{abstract}
$^3$He confined in aerogel in the millikelvin temperature domain exemplifies a Fermi liquid in the presence of disorder. 
In confined $^3$He systems, a solid layer of $^3$He atoms forms on the confining medium.
This system can then be viewed as a model system for the study of the (strongly interacting) Fermi liquid in contact with a (ferromagnetic) "2D-like" adsorbed solid. This interaction, studied experimentally through NMR $T_2$ experiments, is described in the framework of the "fast exchange" model. A {\it complete analytical} descripion of the model is given, explaining our measurements as well as related normal-state confined $^3$He NMR literature.
\end{abstract}

\vspace*{0.2 cm}

\pacs{67.30.-n, 67.30.er, 67.30.hp, 67.30.hr, 68.08.-p, 05.30.Fk}

\maketitle

\section{Introduction}

The nuclear magnetic properties of solid and liquid $^3$He are studied extensively since the 70's, and demonstrate an amazingly rich panel of phenomena: from the ideal (neutral) Fermi liquid, the BCS {\it p}-pairing superfluid, to the magnetic orders U2D2 and CNAF in the solid, associated to their peculiar excitations (a non exhaustive list being particle-hole, spin waves, Homogeneously Precessing Domain, etc) \cite{dobbs,voll}.

Two-dimensional low temperature physics of quantum solids originates in adsorption experiments \cite{dash,vilches,richards}. The study of the magnetic properties of confined $^3$He have followed very rapidly the first results on the bulk liquid \cite{brewer, mylar, carbon, powder, bozler}. It became soon obvious that a few layers of $^3$He were adsorbed on the immersed surfaces, and formed a "2D-like" solid. Especially, with graphite substrates (which present very large surface areas to the adsorbate) ideal 2D magnetic behaviour has been reported: for instance in the Heisenberg ferromagnetic 2D solid, in accordance with the Mermin-Wagner theorem, no phase transition is detected at finite temperature and finite field \cite{osheroff, godfin}. 

With the advent of a new type of porous substrates, namely the silica aerogels, a renewal of the confined $^3$He studies occured in the middle of the 90's \cite{porto,sprague}. An aerogel is a net of strands formed of roughly 3 nm diameter silica spheres. The average distance between strands lies in the range 30-170 nm, which corresponds to samples having porosities lying between 95 \% and 99 \%. Moreover, the structure of the net is fractal over typically two orders of magnitude in lengths. \\
The typical 70 nm size of the aerogel pores makes these samples particularly interesting for $^3$He physics.
Indeed, this lengthscale is of the same order as the superfluid coherence length, which enables to strongly suppress the superfluid transition \cite{porto}. Thus, one can study the effect of a controlled (fractal, with no lattice) disorder introduced in a perfect BCS superfluid. On the other hand, the Fermi liquid properties are not affected \cite{theses,sebtthese}, since their relevant lengthscale is atomic, $k_F^{-1} \approx 1 $ \r{A}. However, the transport properties are strongly modified, since the network of strands limits the mean free path of thermal excitations \cite{saulsus,lancasterheat}. 

In these confined experiments, the behaviour of the liquid and the adsorbed solid at the level of the boundary layer is an intriguing and important question. Indeed, the features observed in Nuclear Magnetic Resonance (NMR, the lineshape in continuous wave, and the T$_1$ and T$_2$ relaxation times in pulsed experiments) are directly linked to the fluid-solid interaction \cite{richardson_shape, hammel_T1, bozler}. Even the state of the matter at the level of the boundary has been questioned \cite{monod}: first of all, is it a liquid or a solid? If it is a liquid, is it a "ferromagnetic liquid"? Or are the two spin baths unchanged, with simply a weak liquid/solid exchange due to the overlap of their wave functions? \\
Moreover, the boundary effects between ferromagnetic and paramagnetic domains is of ubiquitous interests in physics: they appear  here in $^3$He NMR experiments, but can be exploited with ESR (electron spin resonance) \cite{epr} for metallic micro/nano layered structures.

The model of "fast exchange" was rapidly proposed to explain the results obtained on confined $^3$He \cite{richardson_shape,hammel_T1,perry_deconde,richardson}. Based on ideas developed by M.T. B\'eal-Monod and coworkers \cite{monod}, it explained the distorsion of the superfluid NMR lineshapes \cite{richardson_shape}, the linear in temperature relaxation time $T_1$ \cite{hammel_T1}, and the anomalous $1/T$ dependence of the thermal resistance between the fluid and the cold source (instead of the $1/T^3$ Kapitza resistance) \cite{perry_deconde}. In this model the atoms (carrying a spin) can jump {\it very quickly} from one spin bath to the other (i.e. the solid-like and the liquid ensemble). This generates in turn a spin current at the interface which carries information from one ensemble to the other, producing the signatures described above.\\

In the present experimental work we visualize the fast exchange effect through the $T_2$ (spin-spin relaxation time) measured with cw-NMR on $^3$He confined in aerogel. These original results are corollary to the $T_1$,$T_2$ measurements done on other substrates, but their importance lies in the quality of the data. We present in the first part the experimental facts. In the second part, the $^3$He normal state fast exchange model is given through a complete desciption of the formalism, which is missing in the literature up to now. \\
The aim of the paper is to shed light on the magnetic properties of confined $^3$He, in particular by giving the exact conditions of "fast exchange" (what is meant by {\it very quickly}) and the related parameters. We point out in this article that the fast exchange mechanism can then be used to probe other magnetic properties of the combined liquid/solid system.  

\section{Experiment}

\begin{figure}[h!]
\includegraphics[height=10.cm]{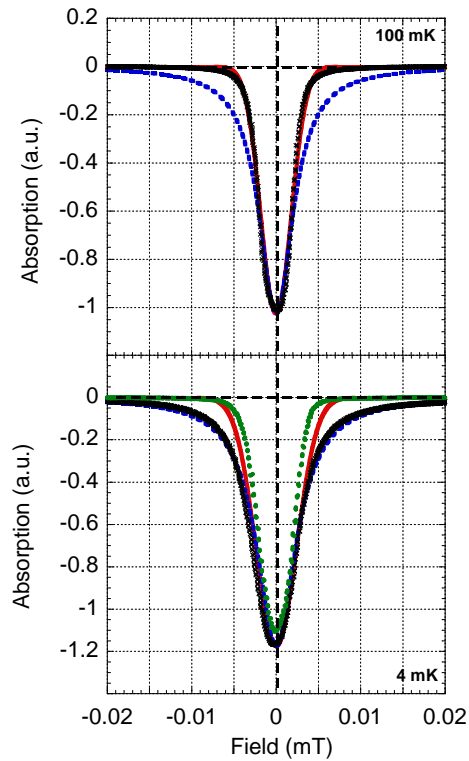}
\caption{\label{fig1} (Color online) $^3$He  cw-NMR (absoption) resonance lines (small crosses) measured at 12.3$\,$bar, at high (100$\,$mK, top) and low (4.1$\,$mK, bottom) temperatures. The zero on the $x$ axis is the $B_0$ applied field (37$\,$mT). The dashed lines (blue) are Lorentzian fits, while the full lines (red) are Gaussian fits. Typically, the low temperature line is Lorentzian, and broader than the high temperature one. The high temperature line, which reflects the field inhomogeneity, looks Gaussian with an inhomogeneous broadening of the order of $4.5\,\mu$T. In the bottom graph the lineshape of the pure liquid, obtained when $^4$He is added, is displayed for comparison (dots, 17$\,$bars at same temperature, green online). }
\end{figure}

In the present paper we report on continuous wave Nuclear Magnetic
Resonance experiments (cw-NMR) performed on $^3$He confined in
aerogel, for pressures ranging from 0 to 30 bars. 
We have used a standard cylindrical 98\% aerogel \cite{mulders}.
The aerogel was inserted in a 5 mm diameter cylindrical cell.
The gap between the wall of the cell and the aerogel was made to be
about 0.1 mm. The pick-up NMR saddle-coils were mounted slightly higher than the closed
bottom of the cell. The upper end of the aerogel sample was about 10 mm
above the coils sensitivity region. An important issue in our experiments
is the homogeneity of the static magnetic field $B_0$ applied vertically, parallel to the aerogel sample (37$\,$mT).
The field distribution gets convoluted to the actual NMR resonance line, giving rise to an "inhomogeneous linewidth" $\Delta B_{inh}$  and an "inhomogeneous lineshape". We  achieved a $4.5\,\mu$T $\Delta B_{inh}$ (full width measured at half height of the absorption, equivalently $145\,$Hz in the frequency domain, see Fig. \ref{fig1} and discussion below). 
 
\vspace{0.2cm}

Two vibrating wire resonators especially calibrated, mounted above the aerogel sample, were used to determine
accurately the $^3$He temperature between about $1\,$mK and a $120\,$mK \cite{VWR}. 

\begin{figure}
\vspace{0.2cm}
\includegraphics[height=6.7cm]{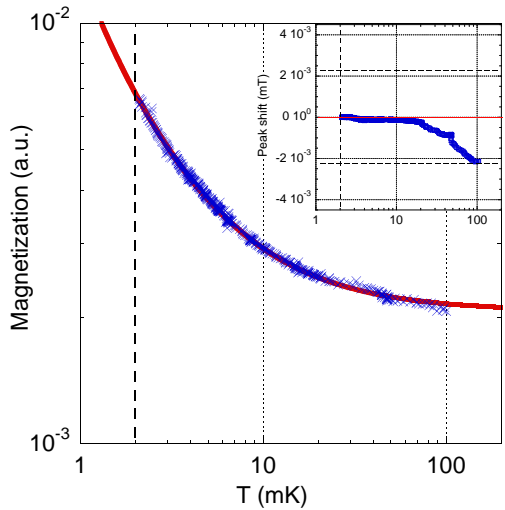}
\caption{\label{fig2} (Color online) Magnetization (area) of the NMR absorption line as a function of temperature, at 12.3$\,$bar and 37$\,$mT. The flat high temperature end is characteristic of the Fermi liquid, while the low temperature growth is characteristic of the adsorbed solid. The line is a fit (see text). Inset: peak position of the line as a function of the temperature; the horizontal dashed lines represent the full-width at half height while the full line is the average resonance field retained. Note the field resolution on the $y$ scale. On both graphs the vertical dashes represent the bulk $^3$He superfluid $T_c$. }
\end{figure}

\subsection{Magnetic properties}

$^3$He gets adsorbed on the silica strands and forms a disordered "2D-like" solid. When necessary, it can be removed by adding controlled amounts of $^4$He (non-magnetic) to the system: due to its larger mass, it adsorbs preferentially and replaces the solid $^3$He.

The characteristics of the cw-NMR absorption resonance line were studied as a function of pressure and temperature, in small radio-frequency drives: namely its area (corresponding to the magnetization), its position (corresponding to the local magnetic field) and its width. The width  
is a function of both the field inhomogeneity $\Delta B_{inh}$, and the intrinsic spin-spin relaxation rate of the system $1/T_2$.

\begin{figure}
\vspace{0.2cm}
\includegraphics[height=6.7cm]{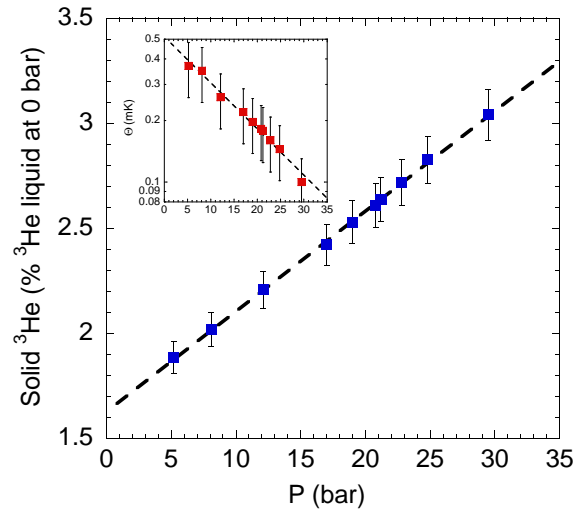}
\caption{\label{fig1bis} (Color online) Amount of solid adsorbed (normalized to the liquid quantity at 0$~$bar) and Curie-Weiss temperature (inset). The monotonous increase of the solid fraction and the decrease of the interaction (seen through $\Theta$) is characteristic of the growth and densification of the disordered solid.}
\end{figure}

In Fig. \ref{fig2} we plot the magnetization $M$ extracted from the NMR absorption line and its position (inset). At low temperature, the magnetization grows almost as $1/T$, which is characteristic of the adsorbed solid. At high temperature, the magnetization flattens out: the solid magnetization is negligible and we recover the Fermi liquid (Pauli) magnetization. It can be fitted to a coexistence of a solid plus a liquid in weak interaction:
\begin{eqnarray*}
M &= & M_{l}+M_{s} \\
M_{l} &=& C_{0}\, \frac{n_{liq}(P)}{T_{F}^{**}(P)} \\
M_{s}&= &C_{0}\, \frac{n_{sol}(P)}{T-\Theta(P)}
\end{eqnarray*}
where $\Theta$ is the Curie-Weiss temperature of the solid (related to the exchange interactions $J$ in the 2D solid and the liquid-solid exchange coupling $I$), $T_{F}^{**}$ the Fermi temperature of the liquid (i.e. interactions in the liquid), $n_{sol}$ and $n_{liq}$ the solid and liquid quantities respectively. $C_{0}$ is the Curie constant per spin. The pressure dependence has been explicitly mentioned. The resulting fitting parameters $n_{sol}(P)/n_{liq}(P=0)$ (in \%) and $\Theta$ are presented in Fig. \ref{fig1bis}, as a function of pressure. 

\begin{figure}
\vspace{0.2cm}
\includegraphics[height=6.7cm]{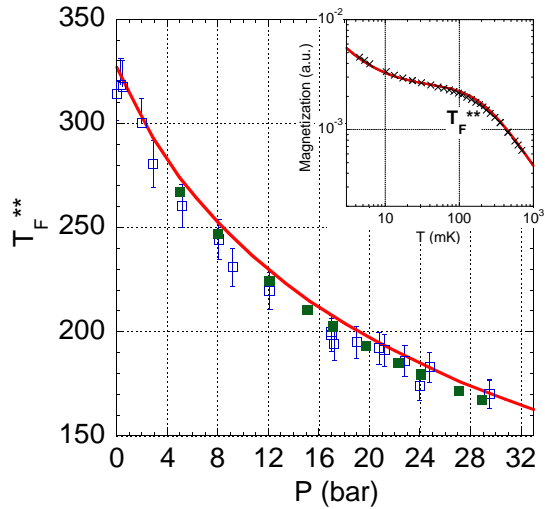}
\caption{\label{figTf} (Color online) Inset: magnetization measurement up to high temperatures realized on liquid $^3$He confined in aerogel (at 17 bar, 25 mT) \cite{sebtthese}. The low temperature growth is the solid contribution already discussed, while the high temperature decrease marks the liquid Fermi temperature $T_F^{**}$ (the line is a guide). The main graph shows the values extracted as a function of pressure (empty squares), in agreement with our results for bulk liquid (full squares, the full line is a guide).}
\end{figure}

The magnetization of the liquid is an important physical parameter of the system. Its magnitude (and thus the strength of the magnetic interactions) is directly given through the effective Fermi temperature $T_{F}^{**}$. In principle, it could be reinforced by exchange with the solid layer \cite{monod}.
We have made detailed measurements of the liquid magnetization, both confined in aerogel and in an open geometry. We find that the $T_{F}^{**}$
 is neither sensitive to the presence of the 100 nanometer-sized disorder, nor to the solid layers (for details, see \cite{sebtthese}). 
This experimental result is given in Fig. \ref{figTf}. \\
The quantity of solid grows linearly with $P$ while the Curie-Weiss temperature decreases, which is characteristic of the densification of a disordered solid \cite{chapellier_sphere}.\\
The inset of Fig. \ref{fig2} shows that at the same time the resonance line position is almost fixed, with a slight drift at high temperatures. This effect is pressure-independent, is also present when the aerogel sample is coated with (non-magnetic) $^4$He, removing the adsorbed $^3$He. It is thus a spurious effect (note the scale in Fig. \ref{fig2}) due to a slightly temperature-dependent magnetic environment. As far as our understanding of the system is concerned, the resonance line position is a constant (the horizontal line of the inset in Fig. \ref{fig2}).  

\subsection{"Fast exchange" on cw-NMR lines}

\begin{figure}
\vspace{0.2cm}
\includegraphics[height=6.5cm]{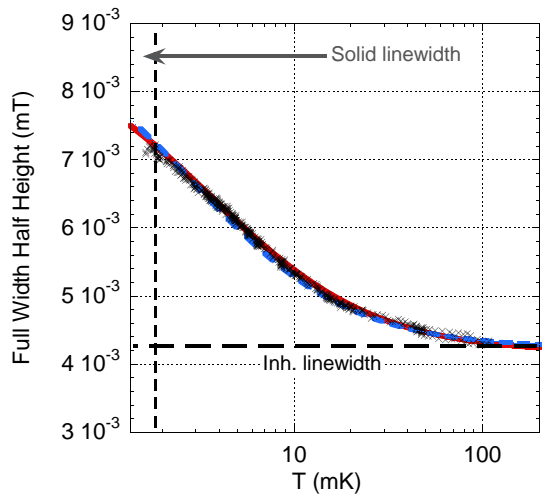}
\caption{\label{fig3} (Color online) Full-width at half height of the NMR absorption line as a function of temperature, at 12.3$\,$bar and 37$\,$mT (crosses). The horizontal dashed line represents the inhomogeneous linewidth, while the arrow at low temperatures represents the linewidth extracted for the solid. The vertical dashed line is the superfluid $^3$He $T_c$. The full line is the fit explained in the text, based on expression (\ref{eq1}). The dashed line is the exact convolution procedure.}
\end{figure}

In Fig. \ref{fig1} we show two typical NMR absorption lines. 
Due to the "fast exchange" of $^3$He atoms, only one common NMR line is seen for the solid plus the liquid components.
At low temperatures the solid dominates, the line looks Lorentzian (a feature of 2D layers \cite{henrireview,henrilineshape}), and is broader than the high temperature one. At high temperatures, we see the inhomogeneous field distribution, which happens to be close to Gaussian. In between, the lineshape changes smoothly, and we can extract the full-width at half height $\Delta B$ as a function of temperature (Fig. \ref{fig3}). The solid linewidth dominates at low temperatures, while its contribution disappears at high temperatures. The field inhomogeneity is understood as a convolution to the liquid-solid NMR line, visualized directly when the aerogel is coated with (non-magnetic) $^4$He (removing thus the solid $^3$He, see Fig. \ref{fig1}). This inhomogeneous linewidth $\Delta B_{inh}$ is much larger than the intrinsic liquid linewidth $1/T_2^{l}$ (see the $T_2^{l}$ reported in the litterature \cite{T2litte}), and is of the order of the intrinsic solid linewidth $1/T_2^{s}$. The fit on Fig. \ref{fig3} is simply a weighted average of the solid and liquid linewidths $\Delta B_{sol}$ and $\Delta B_{liq}$ (including for each the inhomogeneous contribution):
\begin{equation}
\Delta B = \frac{M_{l} \Delta B_{liq} + M_{s} \Delta B_{sol}}{M_{l}+M_{s}} \label{eq1}
\end{equation}
While this procedure neglects the shape difference between Lorentzian and Gaussian lines, the fit is rather good; the exact convolution calculation produces the dashed line. Note that on the contrary it is {\it impossible to fit the data} by simply adding up two (one for the solid and one for the liquid) resonance lines.

\subsection{Resolving the solid}

\begin{figure}[t!]
\vspace{0.2cm}
\includegraphics[height=6.5cm]{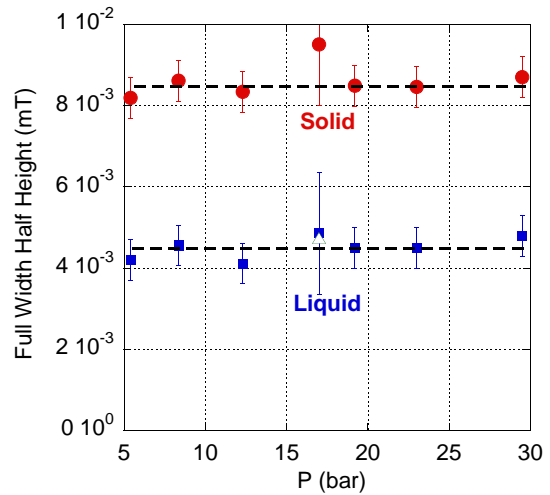}
\caption{\label{fig4} (Color online) Solid and liquid linewidths extracted for all pressures (normal fluid, 37$\,$mT). The liquid contribution directly reflects the inhomogeneous contribution (4.5$\,\mu$T), while the solid term contains the dipolar linewidth narrowed down by exchange couplings, convoluted to the field inhomogeneity. The open symbol is the $^4$He-coated experiment, were no solid contribution could be detected.}
\end{figure}

From (\ref{eq1}) it is possible to extract $\Delta B_{sol}$ and $\Delta B_{liq}$ for various pressures. Both contain the inhomogeneous contribution.
The resulting solid and liquid linewidths are produced in Fig. \ref{fig4}.

The liquid contribution is directly the inhomogeneous contribution $\Delta B_{liq} \approx \Delta B_{inh}$. However, the solid term contains both the true intrinsic solid linewidth and the field inhomogeneity. The intrinsic solid contribution is of order $\Delta B_{sol} \approx 4~\mu$T, which corresponds to a dense solid \cite{henrireview,henrilineshape}. When adding $^4$He, one removes this solid. The pure liquid NMR line reflects then the inhomogeneous field (open symbol, Fig. \ref{fig4}, dots Fig. \ref{fig1}). Moreover, when 20$~$\% of the solid only is left, the lineshape is already the inhomogeneous one, which proves that most of the solid linewidth comes from the first very dense layer \cite{theses}. It explains why the measured solid width seems to be pressure-independent: the first very dense layer is not affected very much by pressurization.

In the following the fast exchange model will be described (within a simplified geometry) in order to explain these linewidth $\Delta B$ (i.e. $T_2$) measurements, together with other NMR confined normal-fluid $^3$He results. The point is that our ability to resolve the solid contribution through the fast exchange formalism makes it a useful tool to study the magnetic properties of the combined system.

\section{Fast exchange model}
\label{introFX}

Nuclear Magnetic Resonance (NMR) is the natural tool used to experimentally access the magnetization of $^3$He systems. It is indeed the technique we used here, and our results have been presented in the previous part. We will therefore expose the following theoretical aspects in the well-known NMR language \cite{abragam,slichter}. The local magnetization will be denoted $\vec{m}(\vec{r},t)$ while the total magnetic moment (the parameter measured in NMR) is $\vec{M}(t)=\int\!\!\!\int\!\!\!\int \vec{m}(\vec{r},t) d^3 r$.

\begin{figure}
\vspace{0.5cm}
\includegraphics[height=5.0cm]{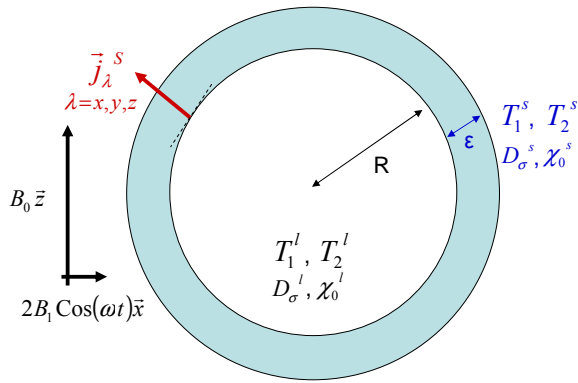}
\caption{\label{fig5} (Color online) Schematic of the two idealized coupled spin baths (not to scale). The global shape is taken to be isotropic for simplicity. The relevant parameters are introduced on the figure, and explained in the text. }
\end{figure}

The model system we consider is a spherical cavity of radius $R$ containing the liquid (in practice about 100$~$nm), in contact at the periphery with a layer of solid of thickness $\epsilon$ (in practice, from one to three "atomic layers", i.e. 1$~$nm). This is schematically represented in Fig. \ref{fig5}. The two spin baths have well defined magnetic relaxation parameters $T_1^{s,l}$ (spin-lattice), $T_2^{s,l}$ (spin-spin) and magnetic transport properties, expressed through a spin diffusion coefficient $D_\sigma^{s,l}$ generating a bulk current $\vec{j}_{\lambda}^{s,l}=- D_\sigma^{s,l} \vec{\nabla} m_{\lambda}^{s,l}$ ($\lambda=x,y,z$ for each magnetization component). The magnetic susceptibility of each spin component is $\chi_0^{s,l}$. A static homogeneous magnetic field $B_0$ is imposed along $\vec{z}$. The two spin baths acquire thus a static (homogeneous) thermodynamical magnetization $m_0^{s,l}=\chi_0^{s,l} B_0$ along $\vec{z}$. The radio-frequency (magnetic) excitation at angular frequency $\omega$ is denoted $2 B_1$ (and points along $\vec{x}$). The total field is thus $\vec{B}=B_0 \vec{z}+2 B_1 \cos (\omega t) \vec{x}$. The properties of both the solid and the liquid are homogeneous. Moreover, the liquid and the solid are supposed to be perfectly isotropic. The superscripts $s,l$ evidently refer to the relevant spin bath. 

In this idealized view, they are two main simplifications which should not be impacting too much the description of the effect we are analysing. First, we wrote "atomic layer" in quotation marks because in practice it certainy {\it is not} a well defined crystalline solid. Moreover, in a 2D solid the intra-"layer" and inter-"layer" spin transports are usually quite different. Nevertheless, we rely on the disordered nature of the solid formed on the porous substrates to somehow "smooth out" these difficulties, by producing an average set of parameters roughly homogeneous and isotropic. The second restriction is that we limit the discussion to one spherical cavity. Again, its properties can be viewed as average parameters obtained over the distribution of pores in the material (aerogel, sinter, powder). What this treatment neglects is any coherent phenomenon coming from the coupling between neigboring cavities (this is seen for instance in one limiting model of a confined superfluid weakly linked from one cavity to the other through a Josephson coupling). In our (normal state) discussion, coherent effects should be negligible. \\
The two spin baths are linked by a boundary magnetic current $\vec{j}_{\lambda}^{S}$ ($\lambda=x,y,z$). The following paragraphs will describe the modeling of the liquid spin bath, of the solid component, and then of this coupling term. \\
In the last part we will solve these equations in simple limiting cases in order to find out the very simple laws the {\it solid+liquid} total system follows in an NMR experiment, making the link with the first experimental part of the paper.

\subsection{Liquid component}
\label{liquidc}

In NMR theory, the lineshape of the absoprtion resonance line of a one family spin system in a homogeneous field is due to the dipolar coupling between the spins. In the paramagnetic solid this linewidth has a Gaussian-like shape and can be quite broad. In the liquid phase however, due to the fast motion of the neighboring particles, the dipolar fields average out and the local field seen by one $^3$He atom reduces almost to the static field $B_0$: the NMR resonance line is very sharp, and this effect is known as {\it motional narrowing} \cite{abragam}. Moreover, the lineshape is almost Lorentzian, which means that the simple Bloch equations will be a very good description of the NMR dynamics. Including the spin diffusion term (due to a bulk current $\vec{j}^{l}_\lambda$ appearing through $D_\sigma^{l}$), they write:
\begin{eqnarray*}
\frac{ \partial \vec{m^l}(\vec{r},t) }{ \partial t}  & = &\gamma \, \vec{m^l}(\vec{r},t) \times \vec{B} - \frac{ m_x^l (\vec{r},t)\, \vec{x}+ m_y^l (\vec{r},t)\, \vec{y}}{T_2^l} \\
 & - & \frac{m_z^l (\vec{r},t) - m_{z0}^l\, \vec{z}}{T_1^l} + D_\sigma^{l} \vec{\nabla}^2 \vec{m^l}(\vec{r},t) 
\end{eqnarray*}
By definition, at the boundary between liquid and solid we have:
\begin{displaymath}
 - D_\sigma^{l} \vec{\nabla} m_{\lambda}^{l}(\left|\, \vec{r}\,  \right| =R,t) = \vec{j}^{S}_\lambda (\left|\, \vec{r}\,  \right| =R^-)
\end{displaymath}
with the notation $R^-$ meaning "on the internal side of the boundary spherical surface". This surface current will be discussed explicitly below.

One important result due to motional narrowing is that (for low enough fields) one simply has $T_1^l=T_2^l$ \cite{bloemb,abragam}. 
In the degenerate Fermi liquid, only one parameter governs both the relaxation times and the spin diffusion coefficient $D_\sigma^{l}$: the quasi-particle scattering time $\tau$. This single particle lifetime scales as $1/T^2$ typically below 50$~$mK \cite{voll,dobbs}. One has simply $T_1^l \propto D_{\sigma}^{l}$. \\
The relaxation time gets much longer than a 100$~$s at low temperatures \cite{T2litte,dobbs}, and the diffusion coefficient has values far above $10^{-5}~$cm$^2$/s \cite{diff,dobbs} (this minimum occuring for both around 0.5$~$K). These values depend on pressure, and are the smallest at the melting curve. This means that on the scale of the solid linewidth, the liquid resonance line is almost a delta function, and the high value of the spin diffusion coefficient will ensure that magnetization is easily transported over the liquid sphere.\\

Due to the high-symmetry of the problem, each $\lambda=x,y,z$ component of the magnetization depends only on $r=\left| \, \vec{r}\,  \right|$. Thus the derivation operators written above reduce to simple expressions, and $\vec{j}^{S}_\lambda \propto \hat{r}$ is  constant over the boundary surface. \\
The above equations describe a trivial precession at $\omega$ of the magnetization about the field axis $\vec{z}$, plus the motion induced by the excitation $B_1$. It is convenient to transpose them in a rotating frame, rotating at the precession velocity, in order to deal only with the slow dynamics induced by the NMR protocol:
\begin{eqnarray*}
\frac{ \partial \tilde{m_x^l}(r,t) }{ \partial t}  & = & - \frac{ \tilde{m_x^l}(r,t) }{T_2^l} + \Delta \omega \,\, \tilde{m_y^l}(r,t) \\
 &+& D_\sigma^{l} \nabla^2 \tilde{m_x^l}(r,t) \\
\frac{ \partial \tilde{m_y^l}(r,t) }{ \partial t}  & = & - \frac{ \tilde{m_y^l}(r,t) }{T_2^l} - \Delta \omega \,\, \tilde{m_x^l}(r,t) \\
&+& D_\sigma^{l} \nabla^2 \tilde{m_y^l}(r,t) -\omega_1 \tilde{m_z^l}(r,t)\\
\frac{ \partial \tilde{\delta m_z^l}(r,t) }{ \partial t}  & = & - \frac{ \tilde{\delta m_z^l}(r,t) }{T_1^l}  \\
&+& D_\sigma^{l} \nabla^2 \tilde{\delta m_z^l}(r,t) +\omega_1 \tilde{m_y^l}(r,t)
\end{eqnarray*}
where the tilded parameters are rotating frame transformed parameters. We have introduced $\Delta \omega = \omega - \omega_0$ with $\frac{\omega_0}{2 \pi}= - \gamma B_0$ and $\frac{\omega_1}{ 2 \pi} = - \gamma B_1$. The quantity $\delta m_z^l(r,t)=m_z^l(r,t) - m_{z0}^l$ is the deviation of the $z$ component from the thermodynamic equilibrium. Note that the $z-$component is not affected by the rotating frame transformation, and the tilde notation can be equivalently used or omitted. \\

If the NMR drive $B_1$ remains small, linear response theory can be applied. The signal measured by the NMR pick-up coil is then $\tilde{M}_t^l e^{i \omega t}$ with $\tilde{M}_t^l=\tilde{M}_x^l + i \tilde{M}_y^l$ (written in complex form, $M$ being the total magnetic moment present inside the coil). The real part of $\tilde{M}_t^l$ is thus proportional to the dispersion $\chi'$ of the AC susceptibility, while the imaginary part is the absorption $\chi''$. 
Without spin diffusion (and in an homogeneous field), the width at half height of the Lorentzian absorption resonance curve is given by $\Delta B_{liq} = 1/(\pi T_2^l \, \gamma)$ in magnetic field units ($\gamma$ is the gyromagnetic ratio, in Hz/T). \\
To compute the actual NMR lineshapes, one thus needs to resolve the set of coupled equations:
\begin{eqnarray}
\frac{ \partial \tilde{m_t^l}(r,t) }{ \partial t}  & = & -(\frac{1}{T_2^l} +i\,  \Delta \omega )\,  \tilde{m_t^l}(r,t)  \nonumber \\
 &+& D_\sigma^{l} \nabla^2 \tilde{m_t^l}(r,t)-i \, \omega_1 \tilde{m_z^l}(r,t) \nonumber \\
\frac{ \partial \tilde{\delta m_z^l}(r,t) }{ \partial t}  & = & - \frac{ \tilde{\delta m_z^l}(r,t) }{T_1^l}  \nonumber \\
&+& D_\sigma^{l} \nabla^2 \tilde{\delta m_z^l}(r,t) +\omega_1 \tilde{m_y^l}(r,t) \label{nmreq} 
\end{eqnarray}
with boundary condition:
\begin{equation} 
- D_\sigma^{l} \vec{\nabla} \tilde{m_{\lambda}^{l}}(r =R,t) = \tilde{\vec{j^{S}_\lambda}} ( r =R^-) \label{bound}
\end{equation}
the spherical symmetry bringing $\vec{\nabla}=\partial/\partial r \, \hat{r}$ and $\nabla^2=1/r \, \partial^2/\partial r^2  \, r$.
The detected liquid signal is obtained by integrating on the cavity volume $4/3 \pi R^3$.

\subsection{Solid component}
\label{solidc}

For a paramagnetic solid, NMR theory predicts a resonance lineshape close to a Gaussian, with a width due to the dipolar coupling $\Delta B_{para} \propto \mu_0 \mu_{^3\!H\!e} /d^3$ ($\mu_0=4\pi . 10^{-7}$ S.I. and $\mu_{^3\!H\!e}$ the $^3$He nuclear magnetic moment, $d$ being the lattice parameter of the solid). Taking the tabulated values and $d \approx 0.5~$nm, one gets $\Delta B_{para} $ of the order of $  100 ~\mu$T. However, if some exchange is allowed in the solid, say a ferromagnetic coupling $J$ between spins (given in Kelvin), then the line is narrowed down essentially for the same reasons as those exposed above for the liquid motion. This is called {\it exchange narrowing}. If this effect is large, the lineshape approaches a Lorentz resonance line, with a linewidth given by $\Delta B_{sol} \propto  \mu_0^2 \mu_{^3\!H\!e}^3 /(d^6 \,J k_B )$. With an exchange $J$ of 10$~\mu$K, the linewidth narrows down to about 1$~\mu$T. The exact values of $\Delta B_{para}$ and $\Delta B_{sol}$ depend on the exact shape of the solid; see the discussion of \cite{abragam} and the original work by Van Vleck \cite{vanvleck}. In the case of a 2D solid, these facts are clearly confirmed experimentally in \cite{henrilineshape}. For our purpose, it means a set of Bloch equations will again be a good description of the magnetization dynamics in NMR experiments. \\
The dipolar field generated by the solid on itself shifts its NMR resonance line \cite{bozlershift}. This shift depends on the orientation of the adsorption surface with respect to the magnetic field $B_0$. As a result, due to the distribution of such orientations in the sample, the NMR solid lineshape broadens and becomes assymetric as the solid polarization increases. In our case, the spherical symmetry minimizes this effect, and the polarization in our range of temperatures is always smaller than 5$~$\%. We can thus safely neglect any solid dipolar broadening or resonance shift, and consider only the case of a perfeclty zero-polarized solid, with a unique (symmetric and Lorentzian) resonance line \cite{theses}.\\

The same equations as those for the liquid (\ref{nmreq}) are valid, replacing the superscript $l \rightarrow s$. The boundary condition replacing (\ref{bound}) writes:
\begin{equation}
- D_\sigma^{s} \vec{\nabla} \tilde{m_{\lambda}^{s}}(r =R,t) = \tilde{\vec{j^{S}_\lambda}} ( r =R^+)  \label{bound2}
\end{equation}
with similar notations to the above ones. 

In the solid, the quantum exchange $J$ is the cause of the spin relaxation $T_1^s$, spin dephasing $T_2^s$ and spin diffusion $D_\sigma^s$. \\
The $T_1^s$ and $T_2^s$ are related to the spectral density of field fluctuations \cite{cowandiff3D} (in the absence of disorder, generated only by the dipolar term, i.e. see the discussion above for the linewidth $\Delta B_{sol}$ giving the inverse of $T_2^s$). Contrary to the liquid case, $T_1^s \neq T_2^s$ . 
For a 2D solid, a careful look at the lineshape (or the free induction decay in pulsed NMR) reveals departures from the simple Lorentzian description \cite{cowan}. It arises from the couplings involved (dipolar and exchange) and the reduced dimentionality. These refinements are outside of the scope of this paper, and average $T_1^s$ and $T_2^s$ will be sufficient to describe the effect discussed here. \\
The spin diffusion coefficient $D_\sigma^s$ can be written in a very general way  $D_\sigma^s \propto J d^2$ \cite{cowandiff3D}. A true (pulsed NMR) spin diffusion experiment is difficult for adsorbed $^3$He, because of the underlying substrate. However, estimates can be obtained from $T_{1,2}^s$ measurements \cite{T2solid}. Typically, values ranging from $10^{-4}~$cm$^2/$s (low density) to $10^{-8}~$cm$^2/$s (high density) are expected. 
From the literature \cite{T2solid,T1solid,japs} one obtains values on the order of 10$~$ms for the $T_1^s$, $T_2^s$ of adsorbed solids in low magnetic fields. 

\subsection{Magnetization current at interface}
\label{magcurrent}

In the problem investigated here, there is no net creation of magnetization at the interface, so the currents on each side of it should be equal:
\begin{displaymath}
\vec{j}^{S}_\lambda ( r =R^-) = \vec{j}^{S}_\lambda ( r =R^+) = j^{S}_\lambda \,\hat{r}
\end{displaymath}
with $\lambda=x,y,z$ for each magnetization component. Due to the symmetry, $\vec{j}^{S}_\lambda$ has to be oriented along $\hat{r}$, and uniform. Moreover, at $r=0$ and $r=R+\epsilon$, the magnetization currents should vanish.

From a microscopic point of view, the current at the interface can be written:
\begin{displaymath}
j^{S}_\lambda  = j^{l \rightarrow s}_\lambda + j^{s \rightarrow l}_\lambda 
\end{displaymath}
with:
\begin{eqnarray*}
j^{l \rightarrow s}_\lambda & = & \frac{1}{\delta S} \sum_{i\,liquid \, \in \delta S} +\Gamma^{l \rightarrow s} \, \mu_{^3\!H\!e} \left\langle \sigma_\lambda^i \right\rangle  \\
j^{s \rightarrow l}_\lambda & = & \frac{1}{\delta S} \sum_{i\,solid \, \in \delta S} -\Gamma^{s \rightarrow l} \, \mu_{^3\!H\!e} \left\langle \sigma_\lambda^i \right\rangle 
\end{eqnarray*}
In the above equations, $\delta S$ represents an infinitesimal element of the boundary surface. On both sides of this surface element (in the {\it liquid} and the {\it solid}) we have a large amount of atoms $i$ denoted by $i \in \delta S$. These atoms contribute to the interface current through the exchange rates $\Gamma^{l \rightarrow s}, \Gamma^{s \rightarrow l}$, with $\mu_{^3\!H\!e} \left\langle \sigma_\lambda^i \right\rangle$ the thermodynamical average of their magnetization ($\sigma_\lambda$ are Pauli operators). Due to the isotropy of the problem, the rates are the same for all directions $\lambda=x,y,z$. The sign arises from the orientation along $\hat{r}$.\\
Introducing $J^{S}_\lambda = \int \!\!\! \int j^{S}_\lambda \, d S= 4 \pi R^2 \, j^{S}_\lambda$, the total surface current can be written:
\begin{equation}
J^{S}_\lambda  =  + C_{liq} \, M_\lambda^l - C_{sol} \, M_\lambda^s \label{current}
\end{equation}
with $C_{sol}$ and $C_{liq}$ two (positive) parameters which are pressure and temperature-dependent (i.e. $C_{sol}=\Gamma^{s \rightarrow l} \,\, n_{sol} \, 4 \pi R^2 $ with $n_{sol}$ the solid contact layer surface density, in at/m$^2$). From the Fermi Golden rule, the exchange rate between a localized spin and the liquid can be written $\Gamma^{s \rightarrow l} = 4 \pi / \hbar \,\, (k_B I)^2 \, N^2(E_F) \, k_B T$ \cite{perry_deconde}. $N(E_F)$ is the density of states at the Fermi level in the $^3$He fluid, and $I$ is the solid-liquid exchange energy (given in Kelvin). \\
Using the same notations as for the magnetizations, we define $j^{S}_t=j^{S}_x+i \, j^{S}_y$ and a tilde denotes rotating-frame transformed currents. \\
The thermodynamical equilibrium of the system imposes $J^{S}_z=0$ when no drive is present ($\omega_1=0$) such that:
\begin{equation}
C_{liq}  =  C_{sol} \, \frac{M_{z0}^s}{M_{z0}^l} \label{equil}
\end{equation}
Note that in this limit, the magnetizations $\vec{m}$ should be homogeneous on each spin bath, and the transverse magnetic current is also necessarily zero $J^{S}_t=0$. \\
By inspecting the above equations, one realizes that the {\it only} interaction parameter which fixes the strength of the exchange is $I$. It is believed that this term is of the order of 100$~$mK \cite{perry_deconde, indirectexch}, which produces $\Gamma^{s \rightarrow l} \approx 1~$MHz at millikelvin temperatures. Furthermore, the impact of even a large $I$ on the solid exchange $J$ is weak, because one needs to involve at least 3 particles (one in the liquid, two in the solid) to modify the solid intra-layer interactions. Typically, contributions to $J$ of the order of 100$~\mu$K are expected \cite{indirecteffectJ,RKKY}.

\section{Solving the equations}

We present below the solution of the above equations in the steady-state case ($\partial \tilde{m_{l,s}^t} / \partial t= \partial \tilde{m_{l,s}^z} / \partial t =0$). \\
In a first part we will give the exact analytical spatial solution of the problem, for low radio-frequency drives $\omega_1 <\!\!< 1/T_2^{l,s},1/T_1^{l,s}$ (the power broadening effects are not discussed). In the second part, we will integrate the equations over the model volume and give the macroscopic NMR properties of the total system.

\subsection{Spatial distribution}

\begin{figure}
\vspace{0.5cm}
\includegraphics[height=4.5cm]{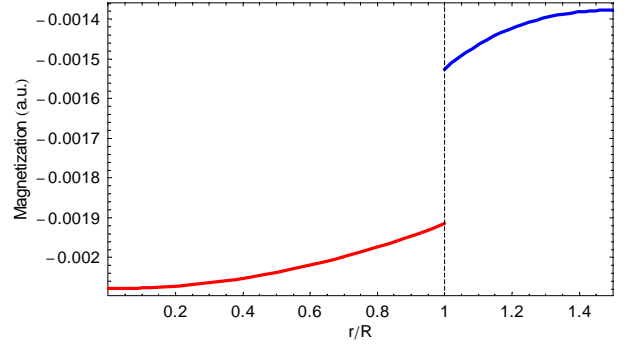}
\caption{\label{maggraph} (Color online) Illustration of the $t$-component (imaginary part) of the magnetizations along $\hat{r}$. The left (red curve) stands for the liquid and the right (blue curve) for the solid; the dashed vertical is the boundary. The parameters chosen for the graph are: $\Delta \omega =0 $, $ T_2^l \, \omega_1 = 0.1$, $ T_2^s \, \omega_1 = 0.001$, $ \epsilon = 0.5   R$, $ C_{sol}=10^{12}~$s$^{-1}$, $ m_{z0}^l=1~$a.u., $ m_{z0}^s=1~$a.u., $\kappa_t^l R=0.1$ and $\kappa_t^s R=1.5$.}
\end{figure}

The fluid components write:
\begin{eqnarray*}
\tilde{m_{t}^l} (r,t)  & = &   - \frac{i\, \omega_1 T_2^l}{1+ i\, \Delta \omega T_2^l} m_{z0}^l  \\
 & - & \frac{(\tilde{j_t^S} R /D_\sigma^l )}{ ( \kappa_t^l R)^2}    \frac{\sinh(\kappa_t^l R \frac{r}{R})}{\frac{\cosh(\kappa_t^l R)}{\kappa_t^l R}-\frac{\sinh(\kappa_t^l R)}{( \kappa_t^l R)^2}}  \frac{R}{r} 
 \end{eqnarray*}
 and:
 \begin{eqnarray*}
\tilde{m_{z}^l} (r,t) & = & m_{z0}^l   \\
 &- & \frac{(\tilde{j_z^S} R / D_\sigma^l)}{ ( \kappa_z^l R)^2}    \frac{\sinh(\kappa_z^l R \frac{r}{R})}{\frac{\cosh(\kappa_z^l R)}{\kappa_z^l R}-\frac{\sinh(\kappa_z^l R)}{( \kappa_z^l R)^2}}  \frac{R}{r} 
\end{eqnarray*}
The solid components are:
\begin{eqnarray*}
     \tilde{m_{t}^s} (r,t)   &   =  &   - \frac{i\, \omega_1 T_2^s}{1+ i\, \Delta \omega T_2^s} m_{z0}^s   \\
   & \!\!\!\!\!\!\!\!\!\!\!\!\!\!\!\!\!\!\!\!\!\!\!\!\!\!\!\!\!\!\!\!\!\!\!\!\!\!\!\!\!\!\!\!\!\!\!\!\!\!\!\!\! -  & \!\!\!\!\!\!\!\!\!\!\!\!\!\!\!\!\!\!\!\!\!\!\!\!\!   \frac{(\tilde{j_t^S} R / D_\sigma^s )}{( \kappa_t^s R+1) \exp(-\kappa_t^s R)\left( \frac{1+\kappa_t^s \epsilon/(\kappa_t^s R+1)}{1+\kappa_t^s \epsilon/(\kappa_t^s R-1)}\exp(-2 \kappa_t^s \epsilon)-1 \right) }  \times \\
   &   & \!\!\!\!\!\!\!\!\!\!\!\!\!\!\!\!\!\!\!\!\!\!\!\!\!\!\!\!\!\!\!\!\!\!\!\!\!\!\!\!  \left( \exp(-2 \kappa_t^s (R+\epsilon) ) \left( \frac{\kappa_t^s (R+\epsilon)+1}{\kappa_t^s (R+\epsilon)-1}\right) \exp(+\kappa_t^s r) + \exp(-\kappa_t^s r) \right)     \\ 
    &  & \!\!\!\!\!\!\!\!\!\!\!\!\!\!\!\!\!\!\!\!\!\!\!\!\!\!\!\!\!\!\!\!\!\!\!\! \times \frac{R}{r}    
 \end{eqnarray*}
 and:
 \begin{eqnarray*}
      \tilde{m_{z}^s} (r,t)     & =   &  m_{z0}^s       \\
  &\!\!\!\!\!\!\!\!\!\!\!\!\!\!\!\!\!\!\!\!\!\!\!\!\!\!\!\!\!\!\!\!\!\!\!\!\!\!\!\!\!\!\!\!\!\!\!\!\!\!\!\!\! - & 
  \!\!\!\!\!\!\!\!\!\!\!\!\!\!\!\!\!\!\!\!\!\!\!\!\!  \frac{(\tilde{j_z^S} R / D_\sigma^s)}{ ( \kappa_z^s R+1) \exp(-\kappa_z^s R)\left( \frac{1+\kappa_z^s \epsilon/(\kappa_z^s R+1)}{1+\kappa_z^s \epsilon/(\kappa_z^s R-1)}\exp(-2 \kappa_z^s \epsilon)-1 \right) }   \times   \\
  &  & \!\!\!\!\!\!\!\!\!\!\!\!\!\!\!\!\!\!\!\!\!\!\!\!\!\!\!\!\!\!\!\!\!\!\!\!\!\!\!\!  \left( \exp(-2 \kappa_z^s (R+\epsilon) ) \left( \frac{\kappa_z^s (R+\epsilon)+1}{\kappa_z^s (R+\epsilon)-1}\right) \exp(+\kappa_z^s r) + \exp(-\kappa_z^s r) \right)      \\ 
    & &  \!\!\!\!\!\!\!\!\!\!\!\!\!\!\!\!\!\!\!\!\!\!\!\!\!\!\!\!\!\!\!\!\!\!\!\!  \times  \frac{R}{r} 
\end{eqnarray*}
all at first order in $\omega_1$, and first order in $\tilde{j_t^S}, \tilde{j_z^S}$.  We introduced the (complex lengths) quantities $\kappa_t^l =\sqrt{\frac{1+i\, \Delta \omega T_2^l}{D_\sigma^l T_2^l}}$, $\kappa_z^l =\sqrt{\frac{1}{D_\sigma^l T_1^l}}$ and $\kappa_t^s =\sqrt{\frac{1+i\, \Delta \omega T_2^s}{D_\sigma^s T_2^s}}$, $\kappa_z^s =\sqrt{\frac{1}{D_\sigma^s T_1^s}}$. The liquid and solid terms are coupled through the (out-of-equilibrium) currents $\tilde{j_t^S}$ and $\tilde{j_z^S}$ which are functions of the drive $\omega_1$. Of course, the above equations reduce to the usual Bloch solutions when $\tilde{j_t^S}=\tilde{j_z^S}=0$. They are illustrated in Fig. \ref{maggraph} with exagerated parameters.\\

The magnetization currents should now be defined self-consistently.
Integrating the above expressions over the sphere for the liquid, and the shell for the solid gives the simple result:
\begin{eqnarray*}
\tilde{M_t^l} & = &  - \frac{i\, \omega_1 T_2^l}{1+ i\, \Delta \omega T_2^l} M_{z0}^l - \frac{ \tilde{J_t^S} \, T_2^l}{1+ i\, \Delta \omega T_2^l}\\
\tilde{M_z^l} & = &  M_{z0}^l - \tilde{J_z^S} \, T_1^l
 \end{eqnarray*}
 and:
 \begin{eqnarray*}
\tilde{M_t^s} & = &   - \frac{i\, \omega_1 T_2^s}{1+ i\, \Delta \omega T_2^s} M_{z0}^s + \frac{ \tilde{J_t^S} \, T_2^s}{1+ i\, \Delta \omega T_2^s} \\
\tilde{M_z^s} & = & M_{z0}^s + \tilde{J_z^S} \, T_1^s
\end{eqnarray*}
Replacing in (\ref{current}) gives finally the expressions for $\tilde{J_t^S},\tilde{J_z^S}$:
\begin{eqnarray*}
\tilde{J_t^S} & = & -i \frac{C_{liq} \, \frac{\omega_1 T_2^l}{1+ i\, \Delta \omega T_2^l} M_{z0}^l - C_{sol} \, \frac{\omega_1 T_2^s}{1+ i\, \Delta \omega T_2^s} M_{z0}^s }{1 +  C_{liq} \, \frac{  T_2^l}{1+ i\, \Delta \omega T_2^l} + C_{sol} \, \frac{  T_2^s}{1+ i\, \Delta \omega T_2^s} } \\
\tilde{J_z^S} & = & o(\omega_1) 
\end{eqnarray*}
The transverse current $\tilde{J_t^S}$ is first order in $\omega_1$ while $\tilde{J_z^S}$ is a second order. Thus, rigorously speaking, there is no spatial dependence of $m_z^{l,s}$ in the first order approach presented in this paragraph (and $T_1^{l,s}$ has dropped out).

From the numerical values quoted in \ref{liquidc} and \ref{solidc} for the spin diffusion $D_{\sigma}^{l,s}$ and the $T_2^{l,s}$ times, we realize that $\left| \kappa_t^l R  \right| <\!\!< 1$ and $\left| \kappa_t^s R \right| \leq  1$. A first order expansion in $\kappa_t^{l,s} R$ of the above expressions is certainly enough to have a good understanding of the phenomena. After simplifications, the final expressions are:
\begin{eqnarray}
\tilde{m_{t}^l}   & = &   - \frac{i\, \omega_1 T_2^l}{1+ i\, \Delta \omega T_2^l} m_{z0}^l  -3 \left( \frac{\tilde{j_t^S}}{R} \right) \frac{T_2^l}{1+ i\, \Delta \omega T_2^l} \nonumber \\
 & & \!\!\!\!\!\!\!\!\!\!\!   \!\!\!\!\!\!\!\!\!\!\! + \left( \frac{\tilde{j_t^S} R}{D_\sigma^l} \right) \left( \frac{3 }{ 10 }  - \frac{1 }{2} \left( \frac{r }{ R}\right)^2 \right) \nonumber \\
m_{z}^l   & \approx & m_{z0}^l \nonumber 
\end{eqnarray}
and:
\begin{eqnarray}
\tilde{m_{t}^s}   & = &   - \frac{i\, \omega_1 T_2^s}{1+ i\, \Delta \omega T_2^s} m_{z0}^s  +  \left( \frac{\tilde{j_t^S}}{\epsilon} \right) \frac{T_2^s}{1+ i\, \Delta \omega T_2^s} \nonumber \\
& & \!\!\!\!\!\!\!\!\!\!\!   \!\!\!\!\!\!\!\!\!\!\! \!\!\!\!\!\! - \left( \frac{\tilde{j_t^S}R}{D_\sigma^s} \right) \left( \frac{1 }{ 2 }   -\frac{2 }{ 3 }\frac{ R }{ r } +\frac{1 }{6} \left( \frac{ r }{ R }\right)^2 +\frac{R }{ \epsilon } \left( \frac{1 }{ 2 } - \frac{ 1 }{ 3 } \frac{ R }{ r } - \frac{1 }{6} \left( \frac{r }{ R}\right)^2 \right) \right) \nonumber \\
m_{z}^s   & \approx & m_{z0}^s \nonumber 
\end{eqnarray}
with:
\begin{eqnarray}
\tilde{j_t^S} & = & - i \frac{(\frac{4}{3} \pi R^3 \,  m_{z0}^l )\, \omega_1}{4 \pi R^2} \nonumber \\
\tilde{j_z^S} & \approx & 0 \nonumber
\end{eqnarray}
where we also took into account $\epsilon <\!\!< R$, $T_2^s <\!\!< T_2^l$ and $C_{sol} T_2^l >\!\!> 1$. This last condition is part of the "fast exchange" limit discussed below. Note that $C_{liq}$ and $C_{sol}$ have disappeared in these last expressions.\\
These results together with Fig. \ref{maggraph} represent our microscopic understanding of the solid-liquid coupled system.
What is expressed by the model is that the magnetization current carries the r.f. excitation from the liquid to the solid, were relaxation occurs (i.e. replace the current expression $\tilde{j_t^S}$ in $\tilde{m_{t}^l},\tilde{m_{t}^s}$ above). Moreover, if the spin diffusion coefficients are large, the magnetization over each spin bath will be practically homogeneous, with a step at $r=R$. In the fast exchange limit, the magnetization current $J_t^S$ is proportional to the drive and to the {\it liquid} magnetization.

\section{NMR properties of total system}
The above section gives an exact view of the magnetization distribution, and the magnetization currents at first order in the driving power $\omega_1$. 
This explicit analytical description is very useful in order to understand the magnetic response of the sample. \\
However, the NMR coil integrates the signal over the cell volume, and in the following we shall deal only with a macroscopic view of the problem. Taking equations (\ref{nmreq}) for the liquid and the solid, integrating over the sphere and the shell volumes respectively, and using Stoke's theorem for the magnetization current:
\begin{eqnarray}
\frac{d \tilde{M_{z}^l}}{d t}   & = & - \frac{1}{T_1^l} \delta \tilde{M_{z}^l} - \tilde{J_z^S} + \omega_1 \tilde{M_{y}^l} \nonumber \\
\frac{d \tilde{M_{z}^s}}{d t}   & = & - \frac{1}{T_1^s} \delta \tilde{M_{z}^s} + \tilde{J_z^S} + \omega_1 \tilde{M_{y}^s} \label{mz}
\end{eqnarray}
and:
\begin{eqnarray}
\frac{d \tilde{M_{t}^l}}{d t}   & = & - \left( \frac{1}{T_2^l} + i \Delta \omega \right) \tilde{M_{t}^l} - \tilde{J_t^S} -i \omega_1 \tilde{M_{z}^l} \nonumber \\
\frac{d \tilde{M_{t}^s}}{d t}   & = & - \left( \frac{1}{T_2^s} + i \Delta \omega \right) \tilde{M_{t}^s} + \tilde{J_t^S} -i \omega_1 \tilde{M_{z}^s} \label{mt}
\end{eqnarray}
the notations have already been introduced. The boundary conditions (\ref{bound}) and (\ref{bound2})  reduce to (\ref{current}):
\begin{displaymath}
J^{S}_\lambda  =  + C_{liq} \, M_\lambda^l - C_{sol} \, M_\lambda^s 
\end{displaymath}
with the equilibrium condition (\ref{equil}):
\begin{displaymath}
C_{liq}  =  C_{sol} \, \frac{M_{z0}^s}{M_{z0}^l} 
\end{displaymath}
In the above equations, the spin diffusion parameters have been integrated out. These equations are fairly general, in particular they hold for any drive $\omega_1$ and any diffusion constants. The only parameter defining the magnetization current is $C_{sol}$. The other terms are thermodynamical parameters of the system $T_{1,2}^{l,s}$ and $M_{z0}^{l,s}$. \\
In the following we will solve the above equations in two simple cases encountered in NMR experiments, within the so-called "fast exchange" limit, that is $C_{liq,sol} \,\, T_{1,2}^{l,s} \,\,\,\, >\!\!> 1$.  

\subsection{T$_1$ measurement}

In a $T_1$ experiment, the magnetization of the system under study is deflected by an r.f. excitation, which is then switched off. During the free induction decay ($\omega_1=0$), the longitudinal component of the magnetization $M_z^{l,s}$ relaxes towards the thermodynamical equilibrium value $M_{z0}^{l,s}$ with an exponential decrease at $T_1^{l,s}$.\\
Here we calculate the spin-lattice relaxation rate of the common NMR resonance line $1/T_1^{avg}$.
Equations (\ref{mz}) can be recast, written in a matrix form:
\begin{displaymath}
\frac{d}{dt} 
\left( 
\begin{array}{c}
      \frac{\delta \tilde{M_z^l}}{M_{z0}^l} \, - \, \frac{\delta \tilde{M_z^s}}{M_{z0}^s}  \\      
   \frac{\delta \tilde{M_z^l}}{M_{z0}^l}+ \frac{\delta \tilde{M_z^s}}{M_{z0}^s}    
\end{array}
\right) 
 =   
\end{displaymath}
\begin{displaymath}
  \!\!\!\!\! \left(\! 
\begin{array}{lr}
       -\left( \frac{1}{2}\left( \frac{1}{T_1^l}+\frac{1}{T_1^s} \right) + (C_{liq}+C_{sol} )\right) \!\!\!\!\!\!\!\!\!\!\!\!\!\!\!\!\!\!\!\!&  -\frac{1}{2}\left( \frac{1}{T_1^l}-\frac{1}{T_1^s} \right)  \\   
   -\frac{1}{2}\left( \frac{1}{T_1^l}-\frac{1}{T_1^s} \right)  \!\!\!\!\!\!\!\!\!\!\!\!\!\!\!\!\!\!\!\!& -\left( \frac{1}{2}\left( \frac{1}{T_1^l}+\frac{1}{T_1^s} \right) + (C_{liq} - C_{sol} )\right) 
\end{array}
\! \right)  
\end{displaymath}
\begin{displaymath}
\times   
 \left( 
\begin{array}{c}
      \frac{\delta \tilde{M_z^l}}{M_{z0}^l} \, - \,  \frac{\delta \tilde{M_z^s}}{M_{z0}^s}  \\      
   \frac{\delta \tilde{M_z^l}}{M_{z0}^l}+ \frac{\delta \tilde{M_z^s}}{M_{z0}^s}       
\end{array}
\right)    
\end{displaymath}
Inspecting the above equations, one realizes that when the conditions $C_{liq} T_{1}^{l} >\!\!> 1, C_{sol} T_{1}^{s} >\!\!> 1$ are fullfilled, the difference of the normalized $z-$component magnetization deflections relaxes quickly to zero. On timescales of the order of $T_1^{l,s}$ one thus has:
\begin{displaymath}
\frac{\delta \tilde{M_z^l}(t)}{M_{z0}^l}  = \frac{\delta \tilde{M_z^s}(t)}{M_{z0}^s}   =  \frac{\delta \left( \tilde{M_z^l}(t)+  \tilde{M_z^s}(t) \right)}{M_{z0}^l+M_{z0}^s}
\end{displaymath}
Adding up equations (\ref{mz}) and injecting this result, one obtains:
\begin{eqnarray*}
\frac{d \left( \tilde{M_{z}^l}+\tilde{M_{z}^s} \right) }{d t}   & = & - \frac{1}{T_1^{avg}}\,  \delta \left( \tilde{M_{z}^l} + \tilde{M_{z}^s} \right)
\end{eqnarray*}
The average relaxation rate $1/T_1^{avg}$ is found to be:
\begin{eqnarray*}
\frac{1}{T_1^{avg}} & = & \frac{\frac{M_{z0}^l}{T_1^l}+\frac{M_{z0}^s}{T_1^s}}{M_{z0}^l + M_{z0}^s} 
\end{eqnarray*}
The magnetization of the total system ($M_{z}^l + M_{z}^s$) relaxes thus with an average rate which is simply {\it the average of the relaxation rates, weighted by the magnetizations}. 
We recover, as we should, the results of \cite{hammel_T1}. \\

One implication of the fast exchange magnetic coupling, linked to the $T_1^s$, is that it enhances the thermal coupling between the liquid and the thermalized solids above the standard Kapitza resistance ($1/T^3$ at low temperatures) \cite{perry_deconde}. In this paper, the authors obtain a very good agreement between the theory of \cite{leggett} and experiments on normal liquid $^3$He. However, we wish to make here a minor comment on this article, which does not affect the results: the authors use the exchange rate $\Gamma^{s \rightarrow l}$ as a measure of the Lorentzian width of the spectral line of the localized spins. This width is in fact not given by $\Gamma^{s \rightarrow l}$, but rather by $T_2^s$ as shown in \ref{solidc}.

\subsection{T$_2$ measurement}

A true $T_2$ measurement can be performed with NMR spin-echo techniques \cite{abragam, slichter}. However, within the field inhomogeneous contribution $\Delta B_{inh}$, an estimate can be obtained through the pulsed NMR free induction decay of the transverse component $\tilde{M_{t}}$, or equivalently the full width at half height of the continuous wave NMR absorption line. \\
Here we caluclate the spin-spin relaxation rate, which is inversely proportional to the intrinsic linewidth of the common NMR resonance line. 
The above equations (\ref{mt}) can be treated as (\ref{mz}), in the case of zero detuning ($\Delta \omega = 0$) and zero r.f. drive ($\omega_1=0$). One obtains symmetrically:
\begin{displaymath}
\frac{  \tilde{M_t^l}(t)}{M_{z0}^l}  = \frac{ \tilde{M_t^s}(t)}{M_{z0}^s}   =  \frac{ \left( \tilde{M_t^l}(t)+  \tilde{M_t^s}(t) \right)}{M_{z0}^l+M_{z0}^s}
\end{displaymath}
within the conditions $C_{liq} T_{2}^{l} >\!\!> 1, C_{sol} T_{2}^{s} >\!\!> 1$.
Adding up (\ref{mt}) and injecting this result brings: 
\begin{eqnarray*}
\frac{d \left( \tilde{M_{t}^l}+\tilde{M_{t}^s} \right) }{d t}   & = & - \frac{1}{T_2^{avg}}\,  \left( \tilde{M_{t}^l} + \tilde{M_{t}^s} \right)
\end{eqnarray*}
with an average relaxation rate for the transverse component:
\begin{eqnarray*}
\frac{1}{T_2^{avg}} & = & \frac{\frac{M_{z0}^l}{T_2^l}+\frac{M_{z0}^s}{T_2^s}}{M_{z0}^l + M_{z0}^s} 
\end{eqnarray*}
This rate is again simply {\it the average of the two baths relaxation rates, weighted by the magnetizations}. 
This was observed by \cite{bozlershift} for $^3$He confined within Grafoil sheets. \\

The continuous wave NMR experiments are corollary to the above results. Take equations (\ref{mt}) in the steady state, with small $\omega_1$ drives. One obtains for the total transverse component $\tilde{M_{t}^l} + \tilde{M_{t}^s}$:
\begin{displaymath}
\!\!\! \tilde{M_{t}^l} + \tilde{M_{t}^s} = i \omega_1 \frac{\alpha_s M_{z0}^l +\alpha_l M_{z0}^s + (C_{liq}+C_{sol} ) (M_{z0}^l + M_{z0}^s) }{-(\alpha_l+C_{liq})(\alpha_s+C_{sol})+C_{liq}C_{sol}}
\end{displaymath}
where we introduced $\alpha_l=1/T_2^l + i \Delta \omega$ and $\alpha_s=1/T_2^s + i \Delta \omega$. \\
When the conditions $C_{liq} T_{2}^{l} >\!\!> 1, C_{sol} T_{2}^{s} >\!\!> 1$ are fulfilled, the above result can be simplified in:
\begin{displaymath}
\tilde{M_{t}^l} + \tilde{M_{t}^s} = - i \omega_1  \frac{M_{z0}^l + M_{z0}^s}{(T_2^{avg})^{-1} + i \Delta \omega}
\end{displaymath}
with $T_2^{avg}$ already introduced. We recover the result that the linewidth in a continuous wave experiment is related to the free induction decay time through $1/\pi T_2^{avg}$ (in the frequency domain). This result is presented in the first experimental part for $^3$He confined in aerogel, Fig. \ref{fig3} and Eq. (\ref{eq1}). Note also that the area of the line is proportional to the total static magnetization $M_{z0}^l + M_{z0}^s$ as it should. In practice, the above results should be convoluted to the field inhomogeneity to quantitatively fit the data (dashed line in Fig. \ref{fig3}).\\

\section{Conclusion}

In the present article we publish experimental NMR (nuclear magnetic resonance) results on normal liquid $^3$He in contact with a ferromagnetic $^3$He solid at very low temperatures. These studies were conducted by immersing a silica aerogel in the fluid. The Fermi liquid properties remain unchanged, while the transport coefficients are limited and a "2D-like" solid forms on the aerogel strands. The importance of our results lies in the quality of the data, enabling a fine study of the liquid/solid coupling, known as "fast exchange". An analytical description of the fast exchange model is given, explaining our data and the related normal fluid literature. \\
The solid and the fluid are in common precession, giving a single continuous wave NMR resonance line, or equivalently a single free induction decay signal. The linear response properties are those of Bloch equations (i.e. Lorentzian resonance) with average parameters $T_{1,2}^{avg}$ obtained from a wheighted average of the two spin baths rates, weighted by the static magnetizations $M_{z0}^{l,s}$. What is clearly stated in the theoretical part is that the fast exchange limit corresponds to $C_{liq,sol} \,\, T_{1,2}^{l,s} \,\,\,\, >\!\!> 1$, with $C_{liq,sol}$ the jumping rates from the liquid to the solid, and vice-versa.\\
The authors want to point out that a thorough understanding of the fast exchange coupling between the solid and the liquid enables to use the effect to probe the magnetic properties of the combined system, especially below the bulk superfluid transition temperature $T_c$. \\

We acknowledge valuable discussions with W. Halperin and A. Andreev. The authors also thank T. Mizusaki and A. Matsubara for their interest in these studies.

\end{document}